\documentclass[12pt,a4paper]{article}
\usepackage[T2A]{fontenc}
\usepackage[utf8]{inputenc}
\usepackage{amsmath,amssymb}
\usepackage{graphicx}
\usepackage{indentfirst}
\usepackage{setspace}
\usepackage{cite}
\usepackage[left=4cm,right=2cm,top=2cm,bottom=2cm,bindingoffset=0cm]{geometry}
\usepackage{mathtext}
\usepackage{mathrsfs} 

\usepackage{color}

\DeclareMathOperator{\err}{err}
\begin{document}

\begin{center}
    {\Large {\bf Effect of strong intersite Coulomb interaction on the topological properties of a superconducting nanowire}}
    \\
    {\large M. S. Shustin, S. V. Aksenov}
    \\
    Kirensky Institute of Physics, Federal Research Center KSC SB RAS, 660036, Krasnoyarsk, Russia
\end{center}

\begin{quote}
For superconducting nanowire with the pairing of extended s-type symmetry, Rashba spin-orbit interaction in a magnetic field, the influence of strong intersite charge correlations on single-particle Majorana excitations is analyzed. This problem is investigated on the basis of the density matrix renormalization group numerical method. It is shown that with an increase in the repulsion intensity of electrons located at the neighboring sites, two subbands emerge in the lower Hubbard band of the open system. Based on calculations of the Majorana polarization and degeneracy of the entanglement spectrum, it was found that a topologically nontrivial phase with one edge state survives at the edge of each of the subbands where the concentration of electrons or holes is minimal. 
\end{quote}

\section{Introduction}
Modern experiments with hybrid superconducting nanowires (SW), characterized by strong spin-orbit
coupling and high g-factor values, do not provide a definite answer to the question of the implementation of Majorana modes (MM) at the boundaries of this structure \cite{chen_19,pan_20}. As a result, there is a need for a more detailed study of the phenomenon of topological superconductivity \cite{elliott_15,valkov_19a,valkov_21} and properties of Majorana states, for example, its nonlocality \cite{deng_18,valkov_19b,aksenov_20b}, spin polarization \cite{sticlet_12,valkov_17a,valkov_17b,valkov_18}, features of the magnetocaloric effect \cite{valkov_17c,valkov_19c}. 

One important issue is the impact of the Coulomb interactions on the Majorana bound states (MBS). Most theoretical studies of these excitations consider quadratic Hamiltonians for which classification of topological superconductivity has been obtained \cite{schnyder_08, kitaev_09}. However, the full screening of the Coulomb interactions in a semiconductor nanowire by a massive superconductor is not generally guaranteed \cite{zaitcev_86}. As a result, a better description of the topological properties of the hybrid nanostructure should include both proximity-induced superconductivity and many-body scattering processes. 

It is worth noting that the action of the gate fields, the number of which can be significant in modern MBS-detection experiments, can considerably influence the intensity of charge correlations in the semiconductor wire. In particular, it was shown in the study \cite{sato_19} that the dependence of current through InAs wires on bias voltage and temperature demonstrates scaling behavior according to the Tomonaga-Luttinger theory (despite the mixing of spin and charge degrees of freedom by strong spin-orbit coupling) \cite{tomonaga_50,luttinger_63,bockrath_99}. The resulting electron-electron interaction parameter indicates the implementation of a strong electron correlation (SEC) regime at low carrier concentrations in the wire.  

One of the factors preventing the MBS detection in the hybrid nanostructure is the suppression of the superconductivity, proximity-induced by the substrate or shell, due to the magnetic field. This problem can be avoided, for example, by assuming that the Cooper pairing in the SW is caused by the coupling with an unconventional superconductor. Then, at a zero magnetic field (preserving time-reversal symmetry) a Majorana Kramers doublet, i.e. the MM pairs at both ends of the wire, emerges \cite{wong_12}. In the case of induced s-wave pairing, such excitations can be obtained by considering two wires with spin-orbit interaction located on opposite surfaces of a conventional superconductor. As a result, the topologically nontrivial phase occurs if the main contribution to the creation of Cooper’s instability comes from the crossed Andreev reflection processes rather than the tunneling of the whole pair into one of the wires. This condition is met precisely in case of strong Hubbard repulsion \cite{thakurathi_18}. It was also noted that in the quasi-one-dimensional SW the relatively high value of the Coulomb repulsion can cause a change of the effective Cooper-pairing potential sign and, consequently, lead to the formation of Kramer’s pairs of MBS \cite{haim_14}. In the SEC regime the occurence of parafermions in systems with the SW was also demonstrated \cite{fradkin_80,fendley_12}.

 It is essential that taking into account strong electron interactions encounters fundamental theoretical problems associated with a significant renormalization of effective interactions, a change in the topological classification, as well as the correct definition and construction of the MM operators \cite{fidkowski_10}. The many-body MM operators were obtained analytically and investigated in detail in the case of the Kitaev chain model \cite{kitaev_01, katsura_15,kells_15}. It was shown that one of the consequences of the existence of the many-body MBS is the stability of the $4\pi$-Josephson effect against to the intersite Coulomb interactions if the Hamiltonian of the whole system (i.e. two tunnel-coupled Kitaev chains) has electron-hole symmetry \cite{boeyens_20}. In addition, in the study \cite{fendley_16} for the spin 1D $XYZ$-model, which can be reduced to the interacting Kitaev chain, the Majorana-type excitation operator was also constructed at a special point of the system parameters.  

The importance of the problem of many-body interactions is also related to the implementation of topological quantum computing. Despite the fundamental stability of the MBS against to local perturbations \cite{ivanov_01}, various processes of phase failure destroying the state of topological qubit still exist. For example, the hybridization of MM wave functions in the short wires or the fluctuations of gate electrostatic potential may be sources of decoherence \cite{zhang_19,lai_18}. In turn, in a number of works, based on the mean-field description or employing the density matrix renormalization group method (DMRG), it was found that both single-site \cite{peng_15} and intersite \cite{ng_15} charge correlations at a certain intensity can increase the resistance of the MBS to different decoherence processes. However, in contrast to these findings, the exact diagonalization of the Hamiltonians of Kitaev short chains shows that accounting of interactions in distant coordination spheres can reduce the lifetime of MBS \cite{wieckowski_19}. 

Using the DMRG method the influence of charge correlations on topological phases is most often considered in the Kitaev chain \cite{turner_11}, as well as the SW model with spin-orbit interaction, which reduces to the Kitaev model in the strong magnetic fields \cite{stoudenmire_11}. The Hamiltonian of such a SW belongs to the D symmetry class where only two phases are possible: a trivial phase with the Majorana number (topological invariant) $\mathscr{M}=+1$ and a nontrivial one with $\mathscr{M}=-1$. The studies \cite{aksenov_20,zlotnikov_20} analyzed the problem of electron-electron interactions in the SW of the BDI symmetry class. The latter means that several topologically nontrivial phases can be implemented: a phase with one MM pair at the end of the structure (similar to the SW of the D class) and a phase with two MBS.   

The above-mentioned studies \cite{aksenov_20,zlotnikov_20} focused on the Hubbard repulsion factor. In this article, the DMRG algorithm examines the effect of strong intersite electron repulsion on the topological properties of the superconducting wire which Hamiltonian belongs to the BDI symmetry class.

\section[Model and method] {Model and method}\label{Sec2}
The SW model under consideration takes into account the Rashba spin-orbit coupling, Coulomb repulsion of electrons, proximity-induced superconductivity and Zeeman splitting. In the tight-binding approximation, the SW Hamiltonian of SW with $N$ sites has the form  \cite{stoudenmire_11, aksenov_20}:
\begin{eqnarray}
\label{Ham_wire}
&~&H  =   \sum_{f=1;\,\sigma}^{N} \Big[\xi_{\sigma} a_{f\sigma}^{+}a_{f \sigma} + \Delta a_{f\uparrow}a_{f\downarrow}+\Delta^{*} a^{+}_{f\downarrow}a^{+}_{f\uparrow}+\frac{U}{2}n_{f\sigma}n_{f\bar{\sigma}}\Big] - \\ &-&\sum_{f=1;\,\sigma}^{N-1} \left( \frac{t}{2}a_{f \sigma}^{\dag} a_{f+1 \sigma} + \frac{\alpha}{2}\eta_{\sigma}a_{f \sigma}^{\dag}a_{f+1 \bar{\sigma}} - \Delta_1 a_{f\sigma}a_{f+1,\bar{\sigma}} - \frac{V}{2}\sum_{\sigma'}n_{f\sigma} n_{f+1,\sigma'}  + \mathrm{h.c.} \right).\nonumber
\end{eqnarray}
Here $t/2$ and $\alpha/2$ are parameters describing the hopping and Rashba spin-orbit interaction between the nearest neighbors, respectively; $\xi_{\sigma}=-\mu+\eta_{\sigma} h$, where $h = \frac{1}{2}g \mu_B H$ -- the Zeeman energy, $g$ -- the Lande g-factor; $\mu_B$ -- the Bohr magneton; $\mu$ -- a chemical potential; $a_{f\sigma} (a^{+}_{f\sigma})$ --  an operator annihilating (creating) electron with a spin projection $\sigma = \uparrow,~\downarrow$ on a site $f$; $\eta_{\uparrow}=1$, $\eta_{\downarrow}=-1$. The quantities $\Delta$ and $\Delta_{1}$ are amplitudes of the proximity-induced superconducting pairing of the extended s-type symmetry. The terms describing on-site, $U$, and intersite, $V$, Coulomb interaction of electrons are also taken into account. The occupation number operator is ${n}_{f\sigma} = a^{+}_{f\sigma}a^{}_{f\sigma}$. Henceforth, we consider all energy variables in units of $t$ and $t = 1$.

Equilibrium properties of the model (\ref{Ham_wire}) in the SEC regime ($U$, $V$ $\gg$ 1) have been studied in the framework of the DMRG approach \cite{white_92b, white_98, white_99}. Its main idea is similar to the other renormalization-group techniques \cite{wilson_75} and consists in the partial exclusion of microscopic degrees of freedom. On the one hand, the resulting effective model acts on some reduced Hilbert space. On the other hand, it describes the main physical properties of the initial one.

The first part of the used DMRG algorith contains the following steps:
\begin{enumerate}
	\item The left cluster $L$ with $N_ {0}$ sites and $ M_ {L}$ eigenstates is considered. The Hamiltonian of such a cluster is given by the formula (\ref {Ham_wire}) as $H_L = H\,|_{N \to N_0 = 3}$. The eigenstates of the Hamiltonian ${H}_{L}$ can be obtained exactly and form a Hilbert space, $\left\{| m_{L} \rangle \right \}$, consisting of $M_ {L} = 64 $ states. Similarly, one can get the Hamiltonian of the right cluster $R$ and construct the corresponding Hilbert space $\left \{| m_ {R} \rangle \right \} $ of its eigen states: $ [\, H_R \,]_{\, \{|\Psi\rangle\} \in \{| m_ {R} \rangle \}} = [\, H_L \,]_{\, \{| \Psi \rangle \} \in \{ | m_{L} \rangle \}}$.
	\item The enlarged cluster S/E (system/environment) is formed by adding the single site to the right/left of the cluster $L/R$. The Hilbert space dimension of the new blocks are $M_{S,E}=M_{L,R} \cdot M_{1}=64\cdot 4 = 256$, where $M_{1}$ -- the single-site Hilbert space dimension. The basis states of the system and environment are acquired as the direct products: $|m_{S}\rangle=|m_{L}\rangle\otimes|m_{1}\rangle$, $|m_{E}\rangle=|m_{1}\rangle\otimes|m_{R}\rangle$. The Hamiltonians of the system and environment are given by 
	\begin{eqnarray*}
    \label{HS_HE}
    H_S &=& H_L\otimes I_1 + I_{N_0}\otimes H_1 +  H_{S,\,int};\nonumber\\
    H_E &=& H_1\otimes I_{N_0} + I_1 \otimes H_R  + H_{E,\,int}.
    \end{eqnarray*}
	Here $I_{N}$ is the identity operator in the space of cluster with $N$ sites. The operator $H_{1}$ ($H_{int}$) can be obtained from the first (second) row of the Hamiltonian (\ref{Ham_wire}) for $N=1$ ($N=2$). The first iteration assumes that $H_{S,\,int}=H_{S,\,int_0}=I_{N_0-1} \otimes H_{int}$, 
	$H_{E,\,int}=H_{E,\,int_0}=H_{int} \otimes I_{N_0-1}$.
	It is essential to note that the adding of single sites in the center of the structure at each iteration step allows to circumvent mock boundary effects \cite{white_92a,schollwock_05}.
	\item The supercluster ($S+E$) with length $2N_{0}+2$ and the Hilbert-space size $M_{S}\cdot M_{E}$ is formed. Its Hamiltonian is
	$$H = H_S \otimes I_{N_0+1}+I_{N_0+1} \otimes H_E + I_{N_0} \otimes H_{int} \otimes I_{N_0}.$$
	The eigenproblem for this Hamiltonian is solved employing the Lanczos algorithm \cite{golub_13}. 
	Here the supercluster Hilbert space is divided into the subspaces with even and odd number of fermions (indexes <<$+$>> и <<$-$>>, respectively):  
	$$H = H_{+} \oplus H_{-};~~H_{\pm}|\,\Psi_{1,2;\pm}\,\rangle=E_{1,2;\pm}|\,\Psi_{1,2;\pm}\,\rangle.$$
	The above expressions explicitly indicate that the Lanczos algorithm is used to find two lowest-energy eigenstates (the ground state and first excited one) for each parity sector. Based on these four quantum states the many-body density operator of the supercluster ($S+E$) is constructed:
	\begin{eqnarray}
    \label{rho_full}
    \rho = \sum_{j=1,2} \Big( p_{j+}|\,\Psi_{j+}\,\rangle\langle\,\Psi_{j+}\,| +p_{j-}|\,\Psi_{j-}\,\rangle\langle\,\Psi_{j-}\,|\Big)
    \end{eqnarray}
	where $p_{1+}+p_{2+}+p_{1-}+p_{2-}=1$. In this work we assume $p_{j+}=p_{j-}=1/4$.

	\item Next, the reduced density matrices are calculated. The corresponding operator of the cluster $S$, $\rho_{S}=Tr_{E}|\psi\rangle \langle\psi|$, can be found by taking the partial trace over the environment degrees of freedom (cluster $E$ and indexes <<$e$>>) \cite{cohen_00}. A similar density operator is formed for the cluster $E$ by tracing over the <<$s$>> indexes. Splitting up these reduced density matrices by fermion parities one can obtain
	\begin{eqnarray}\label{rhoS}
	\rho_{s,s'}=\rho_{s+,s'+}\oplus \rho_{s-,s'-};~~
	\rho_{e,e'}=\rho_{e+,e'+}\oplus \rho_{e-,e'-},
	\end{eqnarray}
	where the explicit form of $\rho_{s\pm, s'\pm}$ is
	$$\rho_{s\pm, s'\pm}=\sum_{u=e_+,\,e_-}\sum_{j=1,2}\langle\, s_{\pm}, u\,|\,\Psi_{j\pm}\,\rangle \langle\, \Psi_{j\pm}\,|\,s'_{\pm}, u\,\rangle,$$
	and similar for the other matrices. They have dimensions $M_{S\left(E\right)}/2 \times M_{S\left(E\right)}/2 = 128\times 128$ for the cluster $S$ ($E$).
    Considering further the cluster $S$, the solutions of the eigenvalue problem for matrices $\rho_{S\pm} \cdot V_{l\pm}=w_{l\pm}\cdot V_{l\pm}$ ($l = 1,\ldots,M_S/2$) determine its effective low-energy degrees of freedom which must be used in the renormalization group procedure. The eigenvalues $w_{l\pm}$ characterize the occupancy of the quantum state $l$ with the fermionic parity $\pm$ and satisfy the condition $\sum_{l} \left(w_{l+}+w_{l-}\right)=1$. Wherein the degeneracy of the eigenvalues means quantum entalglement of the system and environment. Similar remarks are valid for the environment density matrix $\rho_E$. 
	
	\item Thus, the quantum degrees of freedom of the system $S$ are determined by the eigenvectors $V_{l\pm}$ with the largest eigenvalues $w_{l\pm}$. In order to renormilize the Hamiltonians of the $L$ and $R$ clusters the transition matrices $V_{L\pm}=[V_{1\pm},\,V_{2\pm},\ldots, V_{M_L/2;\pm}]$ are built, where $w_{1\pm}\geq w_{2\pm} \geq \ldots \geq w_{M_L/2;\pm}$. The similar matrices are constructed for the cluster $E$. The matrices $V_{L(R)}=[V_{L(R)+},\,V_{L(R)-}]$ have dimensions $M_{S(E)}\times M_{L(R)}$. Finally, the renormilized Hamiltonians and other operators are
	\begin{eqnarray*}\label{DMRG_ren}
    &~&H_{L(R)\pm} = V^+_{L(R)\pm}\cdot H_{S(E)\pm}\cdot V_{L(R)\pm};~~A_{L(R)} = V^+_{S(E)}\cdot A_{S(E)}\cdot V_{S(E)} \nonumber\\
    &~&H_{S,\,int\pm}=\left(V^+_{L} \otimes I_1\right)^+_{\pm}\cdot
    \left(I_1 \otimes H_{S,\,int_0} \right)_{\pm}\cdot\left(V^+_{L} \otimes I_1\right)_{\pm}; \nonumber\\
    &~&H_{E,\,int\pm}=\left(I_1 \otimes V^+_{R} \right)^+_{\pm}\cdot
    \left( H_{E,\,int_0} \otimes I_1 \right)_{\pm}\cdot\left(I_1 \otimes V^+_{R} \right)_{\pm}.
	\end{eqnarray*}
	
	\item The procedure repeats starting from the step 2 until the observables cease to depend on the number of iterations. 
\end{enumerate} 

After the convergence the 1D structure corresponds to the one with a vanishingly small influence of the boundary conditions. This means that the described DMRG algorithm is applicable for infinite systems (infinite DMRG -- iDMRG). Its precision is characterized by a truncation error: $\err = 1 - \sum_{i=1}^{M_L} w_{i}$, where summation is carried out over $M_L$ largest eigenvalues $w_{i}$. In this study calculations $\err \sim 10^{-5}$.  

Next, one has to additionally fix those choices of effective degrees of freedom for the clusters $S$ and $E$ at each iteration step as they are actually made for the superclusters of size lesser than the final length $N$. To do it the finite version of the DMRG procedure is used (finite-system DMRG - fDMRG). In this case the iDMRG procedure continues until the number of sites reaches the value of $N$. Thereafter, the sequence of sweeps is performed in order to achieve the convergence of the algorithm. The idea is to alternately perform the iDMRG procedure only for the left ($L$) or right ($R$) half of the system, keeping the structure length. Simultaneously, the operators of the adjacent ($R$ or $L$) part have to be saved in computer memory and used during the next sweep.  

As a result of the DMRG algorithm, we recieve the approximate many-body quantum states $|\,\Psi_{j\pm}\,\rangle$, energies $E_{j\pm}$ and many-body density matrix (\ref{rho_full}) of the system. The latter allows to study its equilibrium properties. 

\section{\label{Sec3} Results}

In this section the impact of strong local interactions, on-site ($U$) and intersite ($V$) Coulomb repulsion, on the equilibrium properties of the system (\ref{Ham_wire}) is analyzed. During the DMRG numerics we supposed that either $U=5$, $V=0$ or $U=5$, $V=3.5$. It is well known that the strong Hubbard repulsion, $U\gg1$, leads to the splitting of the initial electron band into the lower and upper Hubbard ones \cite{hubbard_63}.

The energy region between them, which is also called the Mott-Hubbard gap, possesses the zero density of states and increases with the $U$ growth. The strong intersite repulsion, $V\gg 1$, causes the additional splitting of each band into two subbands accompanied by the spectral weight redistribution \cite{valkov_11a, valkov_11b}.  

\begin{figure}[h!]
	\begin{center}
		\includegraphics[width=0.5\textwidth]{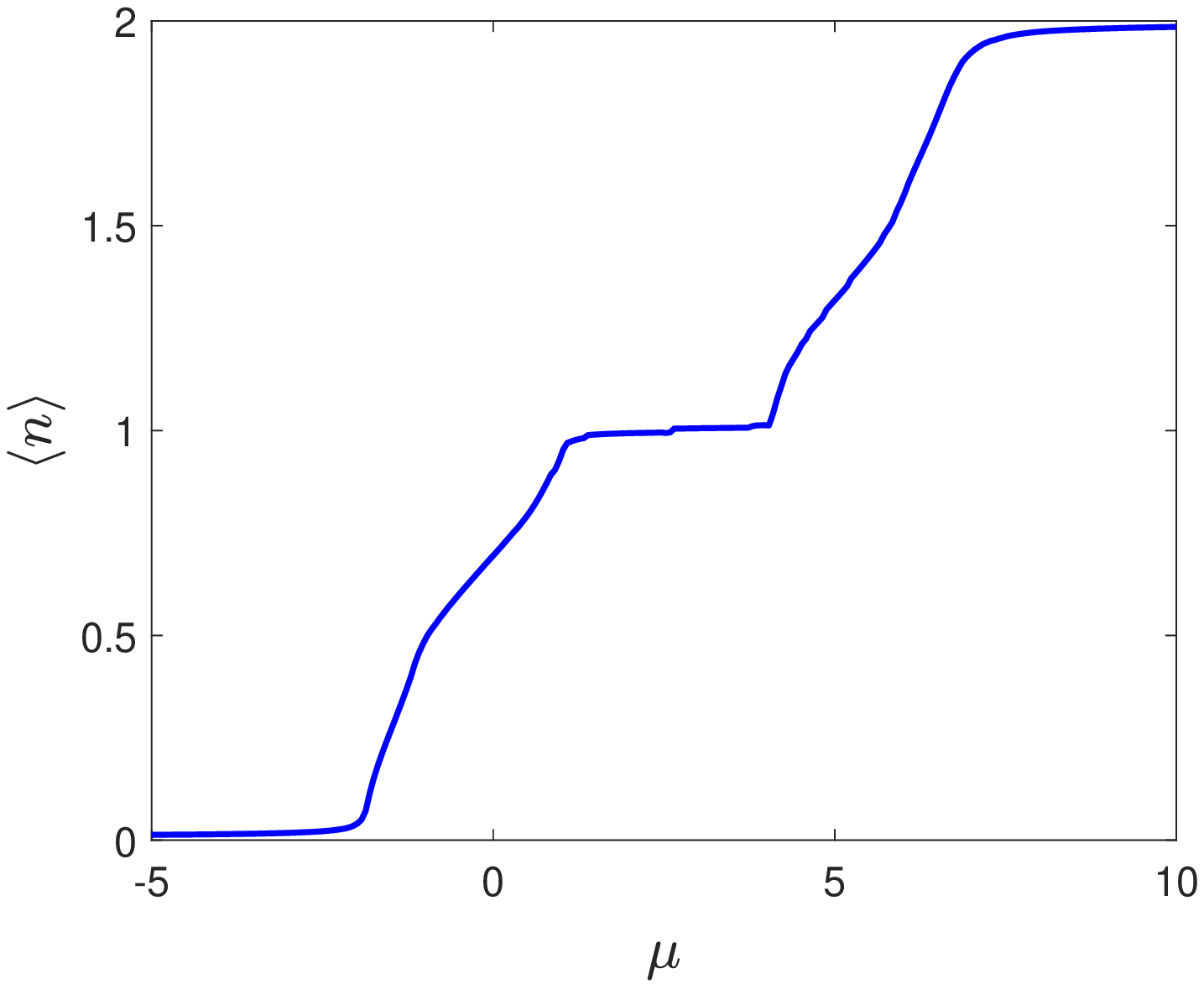}
		\includegraphics[width=0.49\textwidth]{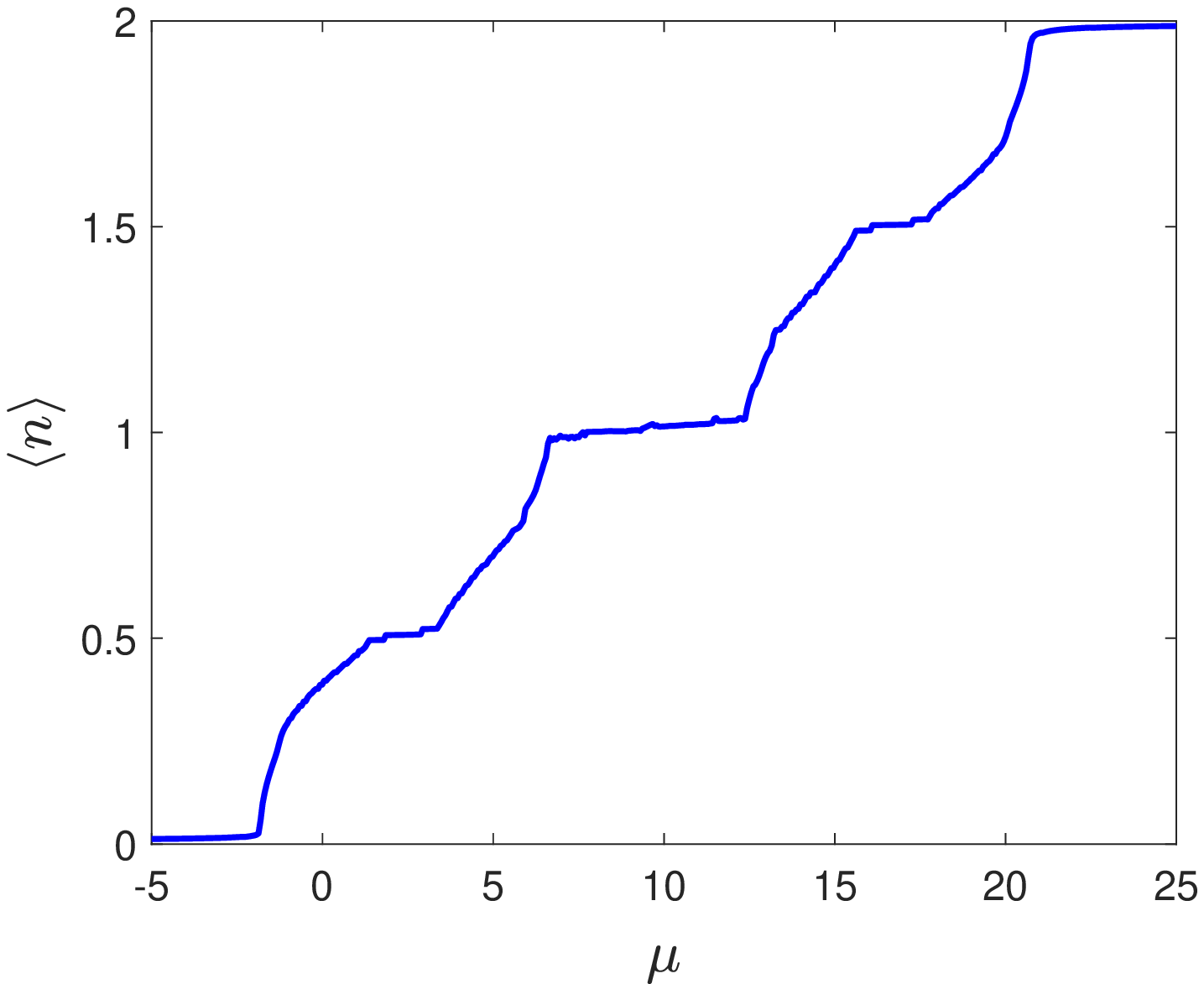}
		\caption{\small The electron concentration vs. chemical potential dependences for strongly correlated superconducting nanowire (\ref{Ham_wire}) in case of zero ($V=0$, (a)) and strong ($V=3.5$, (b)) intersite interaction. The parameters of the model: $t=1$, $\alpha=1.5$, $\Delta=-0.5$, $\Delta=0.2$, $h=0.5$, $U=5$, $N=100$.} \label{nav}
	\end{center}
\end{figure}

For the model (\ref{Ham_wire}) the similar effects, induced by the Hubbard and intersite repulsion, occur. For example, one can observe them in the chemical-potential dependence of the average electron concentration, $\langle\, n \, \rangle = \frac{1}{N} \sum_{f=1}^{N}\langle n_{f} \rangle$. 
The corresponding curves are displayed in Fig.\ref{nav} at $V=0$ (a) and $V=3.5$ (b). If $U=5$, $V=0$ there is the interval $\mu \in [\,1.5\,;~4\,]$ where the $\langle n \rangle\left(\mu\right)$ slope is virtually horizontal caused by the Mott-Hubbard gap. The small nonzero gradient can be attributed to the superconducting pairings which result in the nonzero density of states. Taking into account the intersite interaction leads to the shift of the Mott-Hubbard gap (see the interval $\mu \in [7\,;~13]$ in Fig. \ref{nav}b) and gives rise to the similar gaps inside both Hubbard bands. 

Let us consider in more detail the lower Hubbard band in the SEC regime, $U=5$, $V=3.5$. It emerges in the energy interval $\mu \in [\,-2;~7]$. The subbands induced by the intersite repulsion of electrons are realized at $\mu \in [-2\,;~1]$ and $\mu \in [3.5\,;~7]$.

The topological properties of the lower Hubbard band were analized employing three characteristics: the excitation spectrum, Majorana polarization and entanglement entropy. The former two were found using the fDMRG algorithm. In opposite, the last one were calculated with help of the iDMRG scheme. The lowest excitation energies are defined as:  $\varepsilon_{j} = E_{j-}-E_0$ if $E_{0} = E_{1+}$, and $\varepsilon_{j} = E_{j+}-E_0$ if $E_{0} = E_{1-}$, where $E_0 = \min \{\,E_{1+}\,,\,E_{1-}\}$ - the many-body ground state energy ascribed to the positive and negative fermion-parity sector of the Hilbert space. By the definition $\varepsilon_{j} \geq 0$ and if $\varepsilon_{j} =0$ the quantum phase transition in the system occurs. The Majorana polarization is
\begin{eqnarray}\label{MP}
\textrm{MP}_{j}= \frac{\sum_{f\sigma}' \left( w^2_{jf\sigma} - z^2_{jf\sigma} \right) }{\sum_{f\sigma}' \left( w^2_{jf\sigma} + z^2_{jf\sigma} \right)};~~~~~~
\genfrac{}{}{0pt}{0}
{w_{jf\sigma} = \langle\Psi_j|\left( a_{f\sigma} + a^+_{f\sigma}\right)|\Psi_0\rangle}
{z_{jf\sigma} = \langle\Psi_j|\left( a_{f\sigma} - a^+_{f\sigma}\right)|\Psi_0\rangle;} 
	\end{eqnarray}
where $|\Psi_0\rangle$ is the ground-state wave function; $|\Psi_j\rangle = |\Psi_{j\pm}\rangle$ if $E_0 = E_{\mp}$. Index $'$ nearby the sums means the summation over the half of the system sites. In this study the entanglement entropy is determined in the following way:  
\begin{eqnarray}\label{D}
D = \left[\,\left(N_{it} - N_{s}\right)\cdot \ln(2)\,\right]^{-1}\cdot
\sum_{l = N_s}^{N_{it}}S_{\rho}(l),
	\end{eqnarray}
where $S_{\rho}(l) = \ln\left(d(\rho,\,l)\right)$; $d(\rho,\,l)$ -- a degeneracy of the eigenvalues of the reduced density matrix (\ref{rhoS}); $N_{it}$ -- a number of the iDMRG iterations. We used $N_{it} = 400$, $N_{s} = 300$.   

The above-mentioned quatities allow to distinguish between the different phases. In particular, if the phase is topologically trivial that $\varepsilon_{1,2} > 0$, $\textrm{MP}_{1,2} \to 0$, $D \to 0$. The topologically nontrivial phase with single Majorana bound state is characterized by $\varepsilon_1 \to 0$, $\varepsilon_2 > 0$, $\textrm{MP}_1 \to 1$, $\textrm{MP}_2 \to 0$, $D \to 1$. Finally, if the nontrivial phase has two Majorana bound states that $\varepsilon_{1,2} \to 0$, $\textrm{MP}_{1,2} \to 1$, $D \to 2$. Note that the iDMRG calculation converges well and the $D$ index can be found unambiguously in the strongly interacting system with the moderately suppressed superconducting pairing. Otherwise, the behaviour of $D$ is highly fluctuating (e.g. see Fig.\ref{EMP_V}b).

\begin{figure}[h!]
	\begin{center}
		\includegraphics[width=0.5\textwidth]{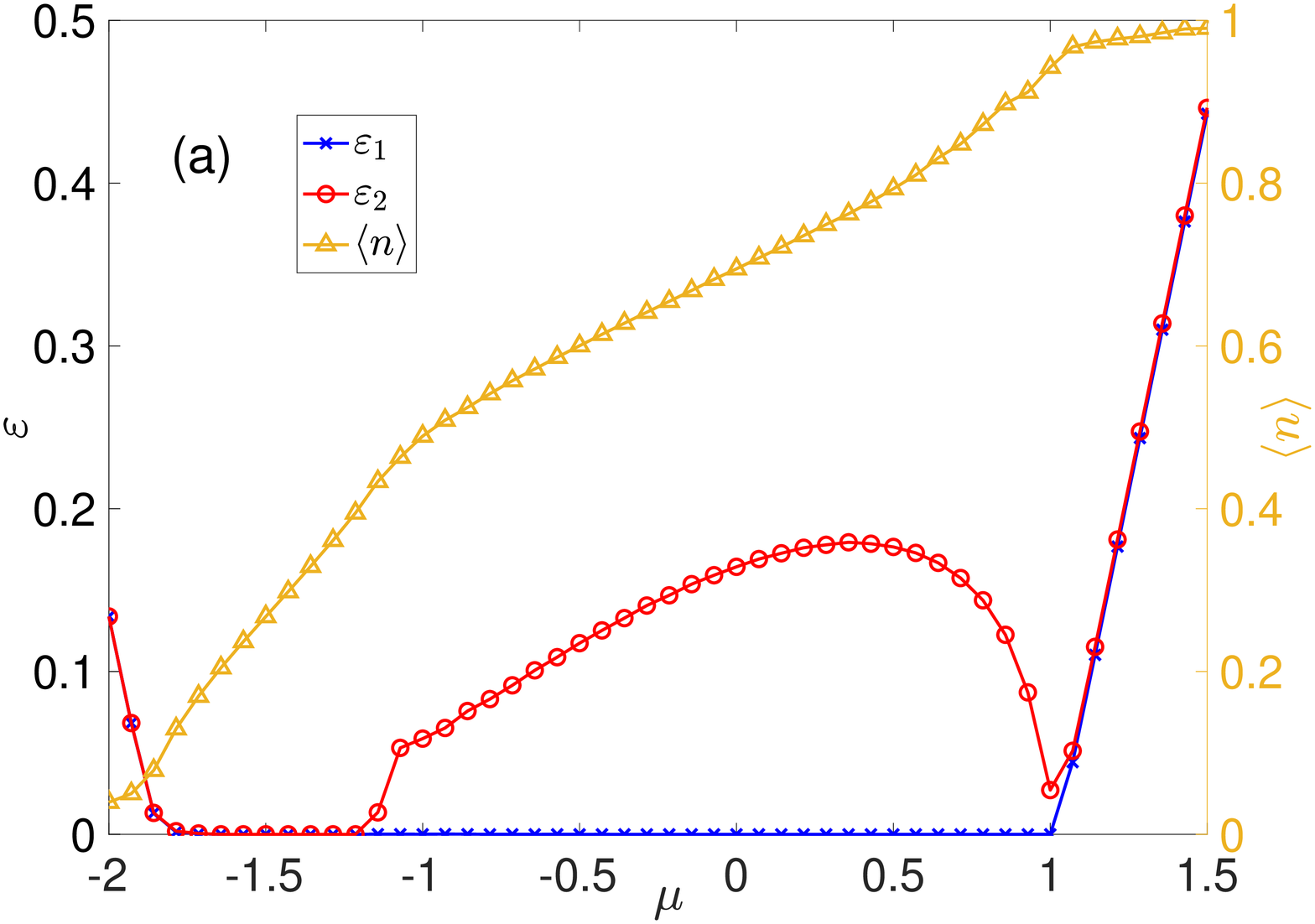}
		\includegraphics[width=0.49\textwidth]{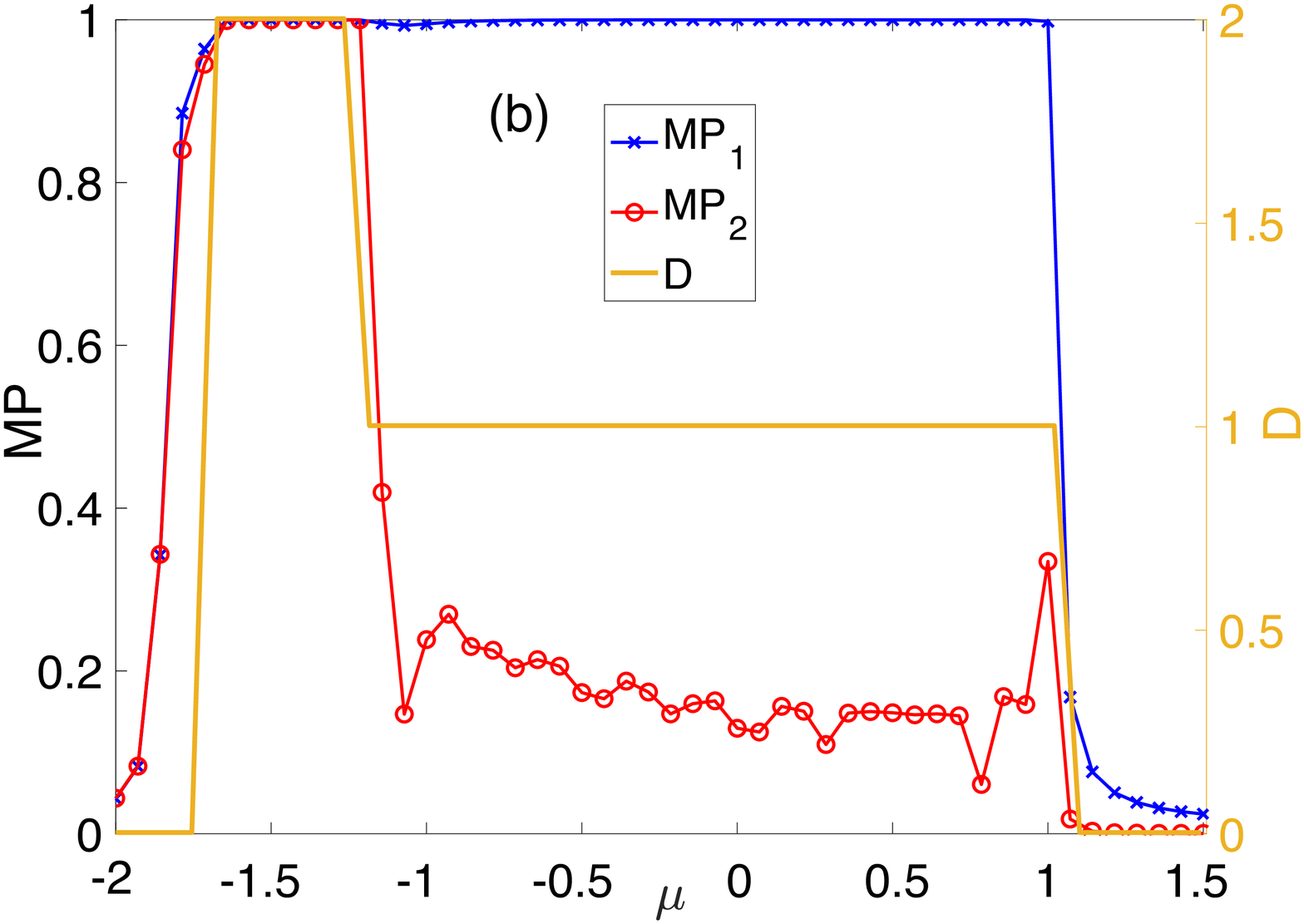}
		\caption{\small The chemical-potential dependence of the energies, average concentration (a), Majorana polarization of the first two excitations of the superconducting wire and the degeneracy of its entanglement spectrum (b) at $V=0$. The parameters are the same as in Fig. \ref{nav}.} \label{EMP_V0}
	\end{center}
\end{figure}

The dependence of two lowest excitation energies on the chemical potential, $\varepsilon_{1,2}\left(\mu\right)$, for the system in the absence of intersite interaction, $V=0$, is shown in Fig.\ref{EMP_V0}a (see the left $y$ axis). The chemical-potential window under consideration comprises the whole lower Hubbard band, i.e. $\langle n\rangle\approx 0$ at $\mu=-2$ and $\langle n\rangle\approx 1$ at $\mu=1.5$ (see the right panel in Fig.\ref{EMP_V0}a). For $\mu>1$ the Mott-Hubbard gap takes place. Inside the lower Hubbard band three different phases show up: 1) a trivial phase at $\mu \in \left[-2;-1.7\right]$; 2) a topological phase with two Majorana bound states at $\mu \in \left[-1.7;-1.2\right]$; 3) a topological phase with single Majorana state at $\mu \in \left[-1.2;1\right]$.
The emergence of the mentioned nontrivial phases is indicated by the behavior of the excitation spectrum, Majorana polarization and $D$ index. The last two quantities are plotted in Fig. \ref{EMP_V0}b. Their $\mu$-dependences point out the edge-like character of the excitations at $\mu \in \left[-1.7;-1.2\right]$ where $\textrm{MP}_{1,2}=1$, $D=2$ and the Majorana nonlocality of the first excitation only at $\mu \in \left[-1.2;1\right]$, i.e. $\textrm{MP}_{1}=1$, $\textrm{MP}_{2}\ll1$, $D=1$. Additionally, a comparison of the $\varepsilon_{1,2}\left(\mu\right)$ and $\langle n \rangle\left(\mu\right)$ dependences implies that the topological transitions are characterized by the peculiarities of the $ \partial \langle n \rangle / \partial \mu $ behavior. Similar features were observed in the Kitaev chain model \cite{chan_2015} and in the model of an electron ensemble on a triangular lattice in the phase of coexistence of superconductivity and magnetism \cite{valkov_2016}. We also note that the analysis of the $\partial \langle n \rangle / \partial \mu $ singularities at $V = 0$ allows one to reconstruct the topological phase diagram of the system obtained in \cite{aksenov_20}.   

\begin{figure}[h!]
	\begin{center}
		\includegraphics[width=0.5\textwidth]{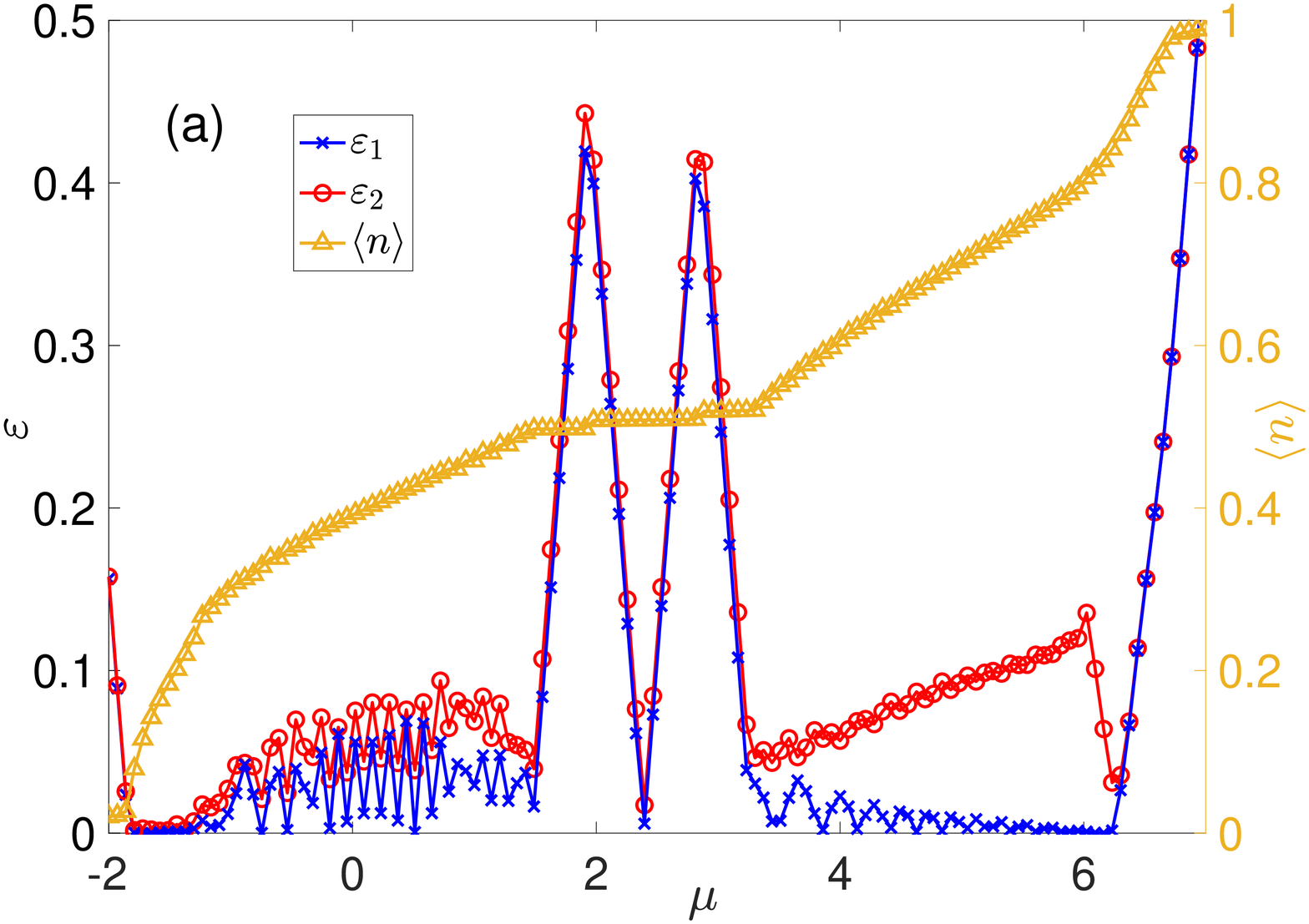}
		\includegraphics[width=0.49\textwidth]{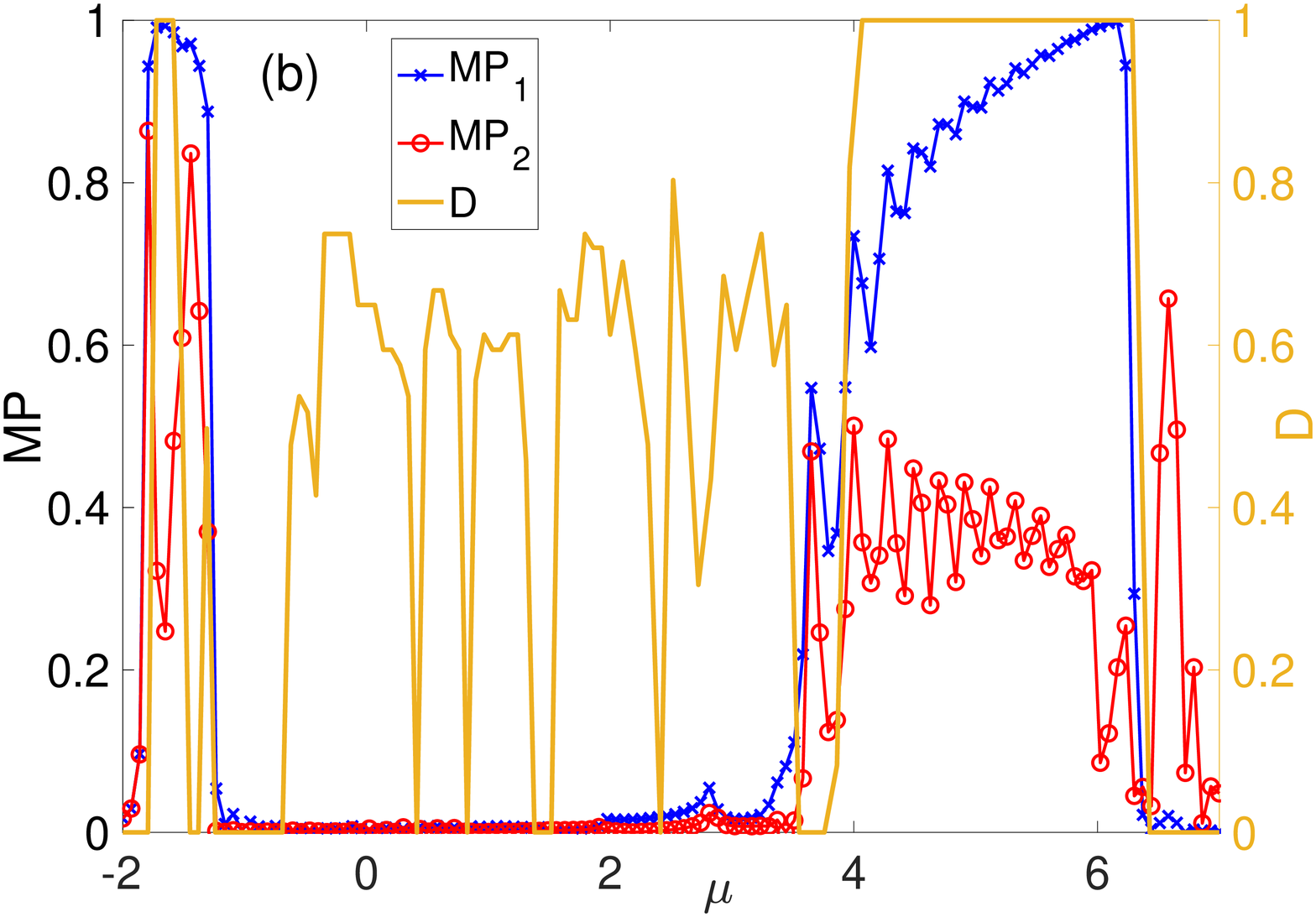}
		\caption{\small The influence of the strong intersite Coulomb correlations on (a) excitation energies and average concentration as well as (b) Majorana polarization and entanglement entropy of the superconducting wire for the parameters of Fig. \ref{EMP_V0} and $V=3.5$.} \label{EMP_V}
	\end{center}
\end{figure}

The influence of the strong intersite repulsion on the excitation energies and electron density in the lower Hubbard band is shown in Fig. \ref{EMP_V}а. One can easily see two subbands established at $\mu \in \left[-2;1.5\right]$ and $\mu \in \left[3;7\right]$ due to these many-body interactions. In the gap the electron concentration reaches a plateau, where $\langle n \rangle \cong 0.5$ (see the right $y$ axis and curve with triangulars). As $V$ grows, the topological phase with single Majorana state begins to be rapidly suppressed for most values of $\mu$. This effect is significantly stronger than the previously discussed in the BDI-class wires which is caused by the strong Hubbard repulsion only \cite{aksenov_20}. It is essential that at $\mu \in \left[-1.9; -1.5 \right]$ and $\mu \in \left[4; 6 \right]$ the realization of single Majorana states is still possible. This observation is also confirmed by the calculations of the Majorana polarization $\textrm{MP}_{1,2}$ and the topological index $D$, which are shown in Fig. \ref{EMP_V}b. The found stability of the topological phases with respect to the strong charge correlations can be explained by the fact that the noted regions of the chemical potential are characterized by a low concentration of electrons or holes, and, accordingly, a relatively weak effect of intersite repulsion on the properties of the system.  

The function $D\left(\mu\right)$ in Fig. \ref{EMP_V}b considerably fluctuates at $V=3.5$ inside the chemical-potential area $\mu \in \left[-1.5;3.5\right]$.  
This behavior is related to the suppression of short-range superconducting pairings and effective breaking of electron-hole symmetry in the system. A similar effect is observed in the case $\Delta_1 = 0$, $V = 0$ and very strong Hubbard repulsion $U>10$. It can be seen that the intersite repulsion leads to a much faster suppression of superconductivity in the system in comparison with the case of taking into account the Hubbard interaction only. 

Finally, we note that when the chemical potential enters the quasi-forbidden bands induced by the Coulomb interactions a quantum phase transition occurs in the system without changing the topological index (see the dependences $\varepsilon_{1,2}(\mu)$ in Fig. \ref{EMP_V}a in the vicinity of $\mu \approx 2.2$). It is important that such effects are due precisely to the presence of intersite repulsion in the system and cannot be found at $V=0$. However, elucidation of the nature and properties of such quantum transitions is beyond the scope of this study. 

\section{\label{Sec5} Conclusions}
The effect of strong intersite electron repulsion on the topological phases of the nanowire with spin-orbit interaction, superconducting pairing of extended s-type symmetry and placed in the external magnetic field is investigated on the basis of the renormalization group method for the density matrix. The analysis performed is a continuation of the study \cite{aksenov_20}, where the superconducting nanowire was characterized by strong Hubbard repulsion, while the intersite Coulomb correlations were supposed to be significantly screened. It is shown that with an increase of the repulsion intensity of electrons located at the neighboring sites the lower Hubbard band splits into two subbands. Between them the region with a very low density of states, similar to the Mott-Hubbard gap, appears. The emergence of two subbands is also observed at the electron concentrations above unity. In each region with a low density of states (both in the Mott-Hubbard and $V$-induced gaps), quantum transitions are observed without a change in the topological index as the chemical potential is swept. Such transitions are realized only in the presence of the sufficiently strong intersite repulsion. 

The topological properties of the described electron subbands in the lower Hubbard band were investigated on the basis of an analysis of the excitation spectrum, Majorana polarization and quantum entanglement spectrum. It is shown that under the strong intersite repulsion the nontrivial topological phases can be found only in the parametric regions characterized by the low concentration of electrons or holes. For the other fillings, the topologically trivial phase with significantly suppressed superconductivity due to the Coulomb interactions is realized. It is noted that the intersite repulsion, as a factor leading to the Cooper pairing destruction, is substantially stronger than the Hubbard interaction.

\section*{\label{Sec5} Financial support} 

This study was supported by the Russian Foundation for Basic Research, the Government of the Krasnoyarsk Territory and the Krasnoyarsk Regional Fund of Science, projects nos. 20-42-243001 and 20-42-243005. M.S.S. thanks the Foundation for the Advancement of Theoretical Physics and Mathematics “BASIS”. S.V.A. is grateful to the Council of the President of the Russian Federation for Support of Young Scientists and Leading Scientific Schools, project No. MK-1641.2020.2

\section*{\label{Sec5} Conflict of interest} 
The authors declare that they have no conflicts of interest.


\begin{thebibliography}{99}

\bibitem{chen_19}
J. Chen et al. Phys. Rev. Lett. {\bf 123}, 107703 (2019).

\bibitem{pan_20}
H. Pan  et al. Phys. Rev. B {\bf 101}, 024506 (2020).

\bibitem{elliott_15}
S.R. Elliott, M. Franz. Rev. Mod. Phys. {\bf 87}, 137 (2015).

\bibitem{valkov_19a}
V.V. Val'kov, V.A. Mitskan, A.O. Zlotnikov et. al. JETP Lett. {\bf 110}, 140 (2019).

\bibitem{valkov_21}
V.V. Val'kov, M.S. Shustin, S.V. Aksenov et. al. Phys. Usp. (2021), accepted for publication DOI: 0.3367/UFNr.2021.03.038950

\bibitem{deng_18}
M. Deng, S. Vaitiekenas, E. Prada et al. Phys. Rev. B. {\bf 98}, 085125 (2018).

\bibitem{valkov_19b}
V.V. Val'kov, M.Yu. Kagan, S.V. Aksenov. J. Phys.: Cond. Matt {\bf 31}, 225301 (2019).

\bibitem{aksenov_20b}
S.V. Aksenov, M.Yu. Kagan. JETP Lett. {\bf 111}, 286 (2020).

\bibitem{sticlet_12}
D. Sticlet, C. Bena, P. Simon. Phys. Rev. Lett. {\bf 108}, 096802 (2012).

\bibitem{valkov_17a}
V.V. Val’kov, S.V. Aksenov. J. Magn. Magn. Mat. {\bf 440}, 112 (2017).

\bibitem{valkov_17b}
V.V. Val'kov, S.V. Aksenov. Low Temp. Phys. {\bf 43}, 546 (2017).

\bibitem{valkov_18}
V.V. Val’kov, S.V. Aksenov. J. Magn. Magn. Mat. {\bf 465}, 88 (2018).

\bibitem{valkov_17c}
V.V. Val'kov, V.A. Mitskan, M.S. Shustin. JETP Lett. {\bf 106}, 798 (2017).

\bibitem{valkov_19c}
V.V. Val'kov, V.A. Mitskan, M.S. Shustin. J. Exp. Theor. Phys. {\bf 129}, 426 (2019).

\bibitem{schnyder_08}
A.P. Schnyder, S. Ryu, A. Furusaki, A. Ludwig. Phys. Rev. B. {\bf 78}, 195125 (2008).

\bibitem{kitaev_09}
A. Kitaev. AIP Conf. Proc. {\bf 1134}, 22 (2009).

\bibitem{zaitcev_86}
R.O. Zaitsev, E.V. Kuz'min, S.G. Ovchinnikov. Sov. Phys. Usp. {\bf 29}, 322 (1986).

\bibitem{sato_19}
Y. Sato et.al. Phys. Rev. B. {\bf 99}, 155304 (2019).

\bibitem{tomonaga_50}
S.I. Tomonaga. Prog. Theor. Phys. {\bf 5}, 544 (1950).

\bibitem{luttinger_63}
J.M. Luttinger. J. Math. Phys. {\bf 4}, 1154 (1963).

\bibitem{bockrath_99}
M. Bockrath et al. Nature. {\bf 397}, 598 (1999).

\bibitem{wong_12}
C. Wong, K. Law. Phys. Rev. B. {\bf 86}, 184516 (2012).

\bibitem{thakurathi_18}
M. Thakurathi et al. Phys. Rev. B. {\bf 97}, 045415 (2018).

\bibitem{haim_14}
A. Haim, A. Keselman, E. Berg, Y. Oreg. Phys. Rev. B. {\bf 89}, 220504(R) (2014).

\bibitem{fradkin_80}
E. Fradkin, L.P. Kadanoff. Nucl. Phys. B. {\bf 170}, 1 (1980).

\bibitem{fendley_12}
P. Fendley. J. Stat. Mech. {\bf 2012}, 11020 (2012).

\bibitem{fidkowski_10}
L. Fidkowski, A. Kitaev. Phys. Rev. B. {\bf 81}, 134509 (2010).

\bibitem{kitaev_01}
A. Y. Kitaev, Phys. Usp. {\bf 44}, 131 (2001).

\bibitem{katsura_15}
H. Katsura, D. Schuricht, M. Takahashi. Phys. Rev. B. {\bf 92}, 115137 (2015).

\bibitem{kells_15}
G. Kells. Phys. Rev. B. {\bf 92}, 081401(R) (2015).

\bibitem{boeyens_20}
J. Boeyens, I. Snyman. Phys. Rev. B. {\bf 102}, 024513 (2020).

\bibitem{fendley_16}
P. Fendley. J. Phys. A. {\bf 49}, 30LT1 (2016).

\bibitem{ivanov_01}
D.A. Ivanov D.A. Phys. Rev. Lett. {\bf 86}, 268 (2001).

\bibitem{zhang_19}
Z.-T. Zhang et al. Phys. Rev. A. {\bf 100}, 012324 (2019).

\bibitem{lai_18}
H.-L. Lai, P.-Y. Yang, Y.-W. Huang, W.-M. Zhang. Phys. Rev. B. {\bf 97}, 054508 (2018).

\bibitem{peng_15}
Y. Peng, F. Pientka, L.I. Glazman, F. von Oppen. Phys. Rev. Lett. {\bf 114}, 106801 (2015). 

\bibitem{ng_15}
H.T. Ng. Sci. Rep. {\bf 5}, 12530 (2015).

\bibitem{wieckowski_19}
A. Wieckowski, A. Ptok. Phys. Rev. B. {\bf 100}, 144510 (2019).

\bibitem{turner_11}
A.M. Turner, F. Pollmann, E. Berg. Phys. Rev. B. {\bf 83}, 075102 (2011).

\bibitem{stoudenmire_11}
E. M. Stoudenmire, J. Alicea, O. A. Starykh, and M. P. A. Fisher. Phys. Rev. B. {\bf 84}, 014503 (2011).

\bibitem{aksenov_20}
S. V. Aksenov, A. O. Zlotnikov, and M. S. Shustin. Phys. Rev. B \textbf{101}, 125431 (2020).

\bibitem{zlotnikov_20}
A. O. Zlotnikov, S.V. Aksenov, M.S. Shustin. Phys. Solid State. \textbf{62}, 1612 (2020).

\bibitem{white_92b}
S.R. White, R.M. Noack. Phys. Rev. Lett. {\bf 68}, 3487 (1992).

\bibitem{white_98}
S.R. White. Phys. Rep. {\bf 301}, 187 (1998).

\bibitem{white_99}
S.R. White, Noack R.M. in: Peschel I., Wang X., Kaulke M., Hallberg K. (Eds.), Density matrix renormalization: a new numerical method in physics. – Berlin: Springer. – 1999.

\bibitem{wilson_75}
Wilson K. Rev. Mod. Phys. {\bf 47}, 773 (1975).

\bibitem{white_92a}
S.R. White. Phys. Rev. Lett. {\bf 69}, 2863 (1992).

\bibitem{schollwock_05}
U. Schollwock. Rev. Mod. Phys. {\bf 77}, 259 (2005).

\bibitem{golub_13}
Golub G., van Loan C.F. Matrix computations. – Baltimore: The Johns Hopkins University Press. – 2013.

\bibitem{cohen_00}
Cohen-Tannoudji C., Diu B., Laloe F. Quantum Mechanics, Vol. 1. WILEY-VCH Verlag GmbH \&
Co. KGaA. – 2020.

\bibitem{hubbard_63}
Hubbard J.C. Proc. R. Soc. London A. {\bf 276}, 238 (1963).

\bibitem{valkov_11a}
Val'kov V. V., Korovushkin M. M. J. Exp. Theor. Phys. {\bf 112}, 108 (2011).

\bibitem{valkov_11b}
Val’kov V., Korovushkin M. J. Phys. Soc. Jpn. {\bf 80}, 014703 (2011).

\bibitem{chan_2015}
Chan Y-H, Sun K. Phys. Rev. B. {\bf 92}, 104514 (2015).

\bibitem{valkov_2016}
Val'kov V.V., Zlotnikov A.O. JETP Lett. {\bf 104}, 483 (2016).

\end{thebibliography}
\end{document}